\documentclass[12pt,preprint]{aastex}

\newcommand{\um}{$\mu$m}
\newcommand{\z}{{\it z}}

\newcommand{\ab}{$\sim$}

\slugcomment{To Appear in the {\it Astrophysical Journal }}

\shorttitle{Supernovae Host Galaxies are Dusty}
\shortauthors{Chary et al.}

\begin{document}


\title{Dust in the Host Galaxies of Supernovae}

\author{R. Chary\altaffilmark{1}, M. E. Dickinson\altaffilmark{2}, H. I. Teplitz\altaffilmark{1}, A. Pope\altaffilmark{3},
S. Ravindranath\altaffilmark{4}}

\altaffiltext{1}{MS220-6 Spitzer Science Center, Caltech, Pasadena CA 91125; rchary@caltech.edu}
\altaffiltext{2}{National Optical Astronomy Observatories, 950 N. Cherry St., Tucson, AZ 85719}
\altaffiltext{3}{Dept. of Physics \& Astronomy, University of British Columbia, Vancouver, B.C. V6T 1Z1, Canada}
\altaffiltext{4}{Space Telescope Science Institute, Baltimore MD 21218}
 
\begin{abstract}

We present Spitzer/MIPS 24$\mu$m observations of 50 supernova host galaxies at $0.1<z<1.7$ in the
Great Observatories Origins Deep Survey (GOODS) fields.
We also discuss the detection of SN host galaxies in SCUBA/850$\mu$m observations of GOODS-N
and Spitzer/Infrared Spectrograph (IRS) 16$\mu$m observations of GOODS-S.
About 60\% of the host galaxies of both Type Ia and core-collapse supernovae are detected
at 24$\mu$m, a detection rate which is a factor of 1.5 higher than the field galaxy population.
Among the 24$\mu$m detected hosts,
80\% have far-infrared luminosities that are comparable to or greater than the optical
luminosity indicating the presence of substantial amounts of
dust in the hosts. The median bolometric luminosity of the
Type Ia SN hosts is $\sim$10$^{10.5}$~L$_{\sun}$, very similar to that of core-collapse SN hosts.
Using the high resolution Hubble/ACS data, we have studied the variation of rest-frame optical/ultraviolet colors within the
24$\mu$m detected galaxies at $z<1$ to understand the origin of the dust emission. The 24$\mu$m detected
galaxies have average colors which are redder by $\sim$0.1 mag than the 24$\mu$m undetected hosts while the latter
show greater scatter in internal colors. This suggests that a smooth distribution of dust
is responsible for the observed mid- and far-infrared emission.
70\% of supernovae that have been detected in the GOODS fields are
located within the half-light radius of the hosts where dust obscuration effects are significant.
Although the dust emission that we detect cannot be translated into a line of
sight A$_{\rm V}$, 
we suggest that the factor of 2-3 larger scatter in the peak $B-V$ colors
that is seen in the high-z Type Ia supernova sample relative to the low-z supernovae might
be partially due to the dust that we detect in the hosts.

\end{abstract}

\keywords{infrared: galaxies -- galaxies: high-redshift -- cosmology: observations}

\section{Introduction}

The use of supernovae (SNe), both Type Ia and Type II, as precision
cosmological tracers is an expanding field with various, dedicated telescope
searches and proposed missions. There are certain possible systematic effects which have been proposed as affecting the calibration
of SNe light curves as distance indicators. Primary among these are grey dust in the intergalactic 
medium (IGM; Aguirre 1999)
or dust in the SN host galaxy. A second systematic is
the proposed variation in the SN light curves with progenitor 
metallicity (e.g. Hoeflich et al. 1998). The evidence for the presence of grey dust
in the IGM is not very strong. It is thought to be heated by the ambient ultraviolet radiation field and be
detectable by a contribution to the far-infrared extragalactic background light.
Since infrared luminous galaxies can account for almost all of 
the DIRBE measured (Hauser et al. 1998) far-infrared background (Chary\&Elbaz 2001; CE), the contribution to the far-infrared
background from dust in the IGM can not be
substantial. Metallicity variations in the progenitors of high redshift SNe are difficult to measure because even large
($\times$3) changes in metallicity result in a change of $\sim$0.3 mag in the $B-V$ colors of the light curve. A decrease
in progenitor metallicity results in observed $B-V$ colors first becoming redder at $z<0.5$ and bluer at high-z. This
color variation cannot be discriminated from the effects of dust reddening until multi-band light curves are created.

The co-moving far-infrared luminosity density of the Universe has been shown to evolve by more than an order of magnitude between
redshifts of 0 and 1 (Chary \& Elbaz 2001). In contrast, the UV luminosity density, increases only by a factor of 2-3 over this redshift range, implying 
that the average dust extinction in the Universe increases with redshift. The bulk of this is 
thought to be due to an evolution in the 
number density of infrared luminous galaxies which emit the dominant fraction of their bolometric luminosity at far-infrared wavelengths and
which are thought to dominate the cosmic star-formation history of the Universe between $0<$z$<2$.
The effect on dust extinction in normal galaxies which dominate the optical/near-infrared
number counts of galaxies is less well known since mid and far-infrared surveys have not
been sensitive to L$_{\rm IR}<$10$^{11}$~L$_{\sun}$ at z$\sim$1. 
The average $B-V$ colors of high redshift supernovae do not show the evolution of dust extinction with redshift inferred from deep 
galaxy surveys (Perlmutter et al. 1999, Riess et al. 2004). This is not a selection effect in SN searches done with the Great Observatories Origins Deep 
Survey (GOODS; Strolger et al. 2004), since the supernovae are detected about 1-2 magnitudes brighter than the limiting sensitivity of
the detection image. 
However, the high redshift color measurement does have a higher intrinsic uncertainty than 
at low redshifts. For example, the 1$\sigma$ $B-V$ color of the SNe at low redshift in the Perlmutter et al. (1999) sample is 0.05 mag
while those at high redshift are around 0.15 mag although the mean of both samples is found to 
be indistinguishable from the SNe at low redshift (see also Riess et al. 2004). 
Attributing the scatter in the colors to variation in the dust extinction along the lines of sight to different supernovae
could imply as much as 0.45 mags of visual extinction
for a Galactic type dust distribution. However, there would have to be an evolution of $\sim$0.15 mag
in the intrinsic peak colors of the SN as well for all the color scatter to be associated with dust.
Grey dust, on the other hand, would be undetectable in $B-V$ color diagrams
and requires optical to near-infrared colors of SNe to be measured (e.g. Nobili et al. 2005).

Type Ia SNe are thought to be end-states in the evolution of 3$<$M$<$8~M$_{\sun}$ mass stars. Since such stars
have a larger contribution to the stellar mass budget and longer evolutionary timescale
than high mass stars, the rates of such SNe in massive galaxies should be 
higher than the rate in low mass galaxies which have not undergone very much star-formation in the past.
Type II+Ib/c SNe (which we collectively refer to as core-collapse or CC SNe)
on the other hand should preferentially be occurring in star-forming galaxies. Since infrared
luminous galaxies dominate the star-formation history of the Universe, core-collapse SNe should mostly have infrared luminous
hosts.
The 24$\mu$m observations of the GOODS (Chary et al. 2005, in preparation) 
undertaken with the Spitzer
Space Telescope (Werner et al. 2004) are sensitive to unprecedented
flux density limits of 10$-$20$\mu$Jy. These observations directly measure the redshifted mid-infrared emission from polycyclic aromatic hydrocarbons and the continuum from very small grains that are transiently heated to T$>$300~K. By adopting the correlation observed between the mid-infrared and far-infrared luminosity of galaxies
in the local Universe it is possible to place constraints on the bolometric luminosity of the SN host galaxies
(Chary \& Elbaz 2001). 

In this paper, we provide mid- and far-infrared
photometry of SNe host galaxies in the GOODS-N
\footnote{http://data.spitzer.caltech.edu/popular/goods/}
and GOODS-S fields (Dickinson et al. 2005, in preparation).
We demonstrate that the SNe host galaxies clearly have significant dust content, despite the fact that the 
average $B-V$ colors from the SNe Ia light curves show little evidence for reddening.
An analysis of the properties of the host galaxy such as their internal colors and color dispersion,
and position of the supernova within the host are also made. We adopt the standard
($\Omega_{\rm M}$=0.27,$\Omega_{\Lambda}$=0.73,H$_{0}$=71 km~s$^{-1}$~Mpc$^{-1}$) cosmology throughout this paper.

\section{Observations}

The 24$\mu$m data presented here were taken as part of the Great Observatories Origins Deep Survey Spitzer Legacy program
(Dickinson et al. 2005; in preparation). A detailed description of the reduction and analysis procedures to obtain the final mosaic and source
catalog will be presented else where (Chary et al. 2005, in preparation). Briefly, the 24$\mu$m observations were taken using the Multiband
Imaging Photometer Spectrometer (MIPS; Rieke et al. 2004). They cover $\sim$165 arcmin$^{2}$ in
each of the GOODS Northern and Southern fields
(center: 12h36m54.9s, 62d14m19.2s and center: 03h32m30.4s, -27d48m16.8s; J2000) up to an exposure time of 10.4 hours per pixel. The data have an angular
resolution of 5.7$\arcsec$ full width half maximum (FWHM) while the relative
astrometric accuracy derived by matching the 24$\mu$m sources with 2MASS stars and the GOODS Infrared
Array Camera (IRAC) images, is better than 0.25$\arcsec$.
All but $\sim$10 low redshift sources are point sources at this spatial resolution. The sources
discussed in this paper are unresolved at MIPS resolution while most are unresolved even at IRAC resolution. 
In the northern field, source catalogs were 
generated by using the positions of sources in the higher resolution IRAC images and fitting groups
of point sources using a singular value decomposition technique at
the positions of the IRAC sources to minimize the effect of source confusion. All SNe hosts and nearby sources which might contaminate 
the 24$\mu$m photometry of the host have accurate IRAC positions which allow all sources in the field to be simultaneously fit for.
This is equivalent to a DAOPHOT-type approach which is commonly adopted to obtain stellar photometry in crowded fields.
The source catalogs
are 84\% complete down to a flux density limit
of 24$\mu$Jy.  
The GOODS-S 24$\mu$m data were made available only in June 2005 and have not undergone the detailed cataloging
effort of the northern field. We have extracted photometry of the host galaxies using the same ``IRAC-prior" technique.
Although the flux uncertainties of the GOODS-S sources
might vary depending on the residual images, we find the photometry in the 
southern field to be reliable at the $\sim$10$-$20\% level.
The Hubble Advanced Camera for Surveys (ACS) $i$ and $z$-band
data which are used in this analysis were taken as part of the Hubble GOODS Treasury 
program and are described in Giavalisco et al. 2004. 

The properties of the SNe in the GOODS fields, including their 
coordinates, redshift and type were compiled from Strolger et al. (2004).
Although we do not see the SN in the ACS mosaics,
some of the SN coordinates have been modified from Strolger et al. (2004). 
The absolute astrometry of the full GOODS/ACS data set enables the position of the host galaxy nucleus
to be estimated to an accuracy of better than 0.25$\arcsec$ and the SN coordinates are re-derived based on the published
offset of the SN from the host.
A few additional SNe (SN1997ff, SN1997fg, SN2002dc, SN2002dd, SN1999gu, SN2003lt, SN2003lu, SN2004R)
were obtained from the IAU CBAT service\footnote{http://cfa-www.harvard.edu/iau/cbat.html} and
were discovered in various past and ongoing searches such as those 
by Gilliland et al. (1999), Blakeslee et al. (2003), Strolger\&Riess (2004), Cappellaro et al. (2000). A listing of 
all the SNe considered in this paper and 24$\mu$m flux densities from their host galaxies
is compiled in Tables 1 and 2.
SN 1997ch, 1997ci and 1997cj, 2001ip and 1999gt although in the vicinity of the GOODS fields, do 
not have 24$\mu$m coverage and are omitted from further discussion.

In order to match the 24$\mu$m data to the high resolution ACS data 
we have firstly, identified the location of the host galaxy of the SNe in the ACS
images from a search of the literature (Strolger et al. 2004; Blakeslee et al. 2003; Riess et al. 2003;
Gilliland et al. 1999).
As we will describe in Chary et al. (2005, in preparation), to alleviate the effect of sources blended in the MIPS 24\um\ images, 
we have utilized the IRAC data which have an intermediate spatial resolution of $\sim$2$\arcsec$
to generate MIPS catalogs. 
Figure 1 shows 
representative snapshots of ACS $z$-band, Spitzer 
IRAC 3.6/4.5 $\mu$m and
MIPS 24$\mu$m images. The Spitzer detections of the SN host galaxies fall under three categories:
\begin{itemize}
\item The supernova host in the IRAC and MIPS images is clearly separated from all nearby galaxies (e.g. SN 1997fg, SN2002dc,
2003N, 2003ba, 2003bb, 2003be, 2003dx, 2003eb, 2003er, 2003et);
\item The supernova host is clearly separated in the IRAC images but the MIPS counterpart is 
blended (e.g. SN 1997ff, 2003eq, 2003es). These have been deblended in our source lists.
\item The supernova host is blended with nearby galaxies in the IRAC images and the MIPS 
images (e.g. SN 2002dd, 2002kh, 2002ki, 2003az, 2003bc, 2003ew). We mostly provide limits for these sources unless it is clear 
that the IRAC counterpart belongs to the host galaxy.
\end{itemize}

Of the 9 host galaxies that are blended at 24$\mu$m at Spitzer resolution,
we are unable to deblend the various components of emission
only for four (SN 2002dd, 2002ki, 2003bc, 2003ew) and we either
provide 1$\sigma$ uncertainties or an upper limit to the total flux density observed at the location of the source.
The hosts of SN2003bd and SN2003fv are unknown/undetected at any optical/infrared wavelength including the new GOODS Spitzer images. 

Flux uncertainties are calculated by the PSF weighted sum of variance values in the 24$\mu$m residual image
after all sources have been fit and subtracted. These are conservative uncertainties which include a term for the
confusion noise whereby the flux density of a nearby source affects the quality of photometry on the source of interest, leaving
a larger residual. 

We have also utilized the SCUBA 850$\mu$m supermap in the HDF-N (Borys et al. 2003)
and Spitzer Infrared Spectrograph (IRS; Houck et al. 2004) 16$\mu$m imaging (Teplitz et al. 2005) of GOODS-S to get additional
constraints on the dust properties of SNe host galaxies. Only one of the SNe hosts in the HDF-N (2002dc) is detected in 
the SCUBA supermap. The presence of a 1.4 GHz radio counterpart allows the 850$\mu$m
source to be securely identified 
with the SN host galaxy (Pope et al. 2005). 

In the GOODS-S, the Spitzer/IRS 16$\mu$m observations reach a limiting sensitivity of $\sim$40$\mu$Jy. Since this is also a 
shorter wavelength compared to the GOODS 24$\mu$m observations, they are effectively a factor of 2$-$4 less sensitive 
than the 24$\mu$m data
for the detection of dusty galaxies. We find that only 5 of the brightest 24\um-detected GOODS-S SN hosts are
detected in the 16$\mu$m survey.
Three of these are CC SNe hosts
(2002kb, 2002hq and 2002fz) while two are the hosts of Type Ia SNe (2002kc and 2002kd). We include these measurements
for completeness since their primary utility is in independently confirming the infrared luminosity derived from the
higher signal-to-noise GOODS 24$\mu$m data.

\section{Origin of Mid-infrared Emission in SNe Host Galaxies}

In local galaxies, the mid-infrared emission arises from very small grains heated in
star-forming regions of the galaxy or in
a nuclear starburst. It has also been shown that the mid-infrared emission can arise from dust
heated by the ambient interstellar radiation field (Li \& Draine 2002).
The mid-infrared luminosity of local galaxies in the IRAS bright galaxy sample (Soifer et al. 1987) has been found to be
correlated strongly with
their far-infrared luminosity which is dominated by large, cool dust grains (CE).
This correlation has been applied to develop a library of model templates for the mid and far-infrared
spectral energy
distribution of galaxies. Given the redshift and observed 24$\mu$m (16$\mu$m) flux density of the source, we can select the template
which gives the closest 24$\mu$m (16$\mu$m) flux density at that redshift to apply a bolometric correction. 
The corrections based on the CE and
Dale\&Helou (1999) templates are used
to derive an infrared luminosity L$_{\rm IR}$=L$(8-1000\micron)$. 
This technique of deriving the bolometric luminosity from the rest-frame mid-infrared luminosity is shown to be accurate
to 40\% in the local Universe (CE01). 
The derived L$_{\rm IR}$ are shown in Figure 2 and
Tables 3 and 4. The difference between the derived L$_{\rm IR}$ from
the two templates is assumed to be representative of the systematic uncertainty in 
the bolometric correction.
Statistical uncertainties are assumed to correspond to the signal to noise ratio of the source at 24$\mu$m.

The fundamental assumption in this technique is that the bolometric correction is dependent only on the
mid-infrared luminosity of the object. In principle, it is possible for sources to have a lower FIR/MIR
ratio for a particular MIR luminosity which would increase the systematic uncertainty in the bolometric 
correction. However, the validity of the mid- to far-infrared correlation 
and the one-to-one correlation between the bolometric correction and the MIR luminosity
has been tested for field galaxy samples out to
\z\ab1 (Appleton et al. 2004, Marcillac et al. 2005) which enables 
the bolometric luminosity to be derived for the SN host sample.

We find that about $\sim$60\% (28/48 excluding SN2003bd and SN2002fv which do not have hosts) of all 
SNe host galaxies are detected in the mid-infrared, and that 
their infrared luminosities span the range of 
typical dusty field galaxies rather than
extreme ultraluminous infrared galaxies. The optical luminosities of these galaxies are calculated as the greater
of $\nu$L$_{\nu}$ in the ACS $i$ or $z$ band.
Among the 24$\mu$m detected host galaxies,
more than 80\% (23/28) have infrared luminosities that are comparable to or
greater than their optical/ultraviolet luminosities. 
Performing a similar analysis with a $0<z<1$ redshift-limited field galaxy sample ($\sim$1700 galaxies)
from GOODS, we find that
36\% of the field galaxies have L$_{\rm IR}$/{L$_{\rm opt}$}$>$0.8, a factor of $>$1.5 lower than the SNe hosts,
indicating the presence of significant quantities of dust within
the SN host galaxies. 

To assess if this difference is because of a difference in the redshift distribution of the two samples of galaxies,
we performed a monte-carlo selection of sources from the field galaxy sample such that the selected sources had a
redshift distribution identical to that of the SN hosts. The fraction of selected
sources whose L$_{\rm IR}$/{L$_{\rm opt}$}$>$0.8 was derived. This selection procedure was repeated a 1000 times. 
We performed a similar procedure such that the selected sources had the same $z$-band apparent magnitude 
distribution as that of the hosts. We find
that in a field galaxy population with the same redshift/apparent magnitude distribution as the SN hosts, the fraction of 
sources with L$_{\rm IR}$/{L$_{\rm opt}$}$>$0.8 is only 32$\pm$8\%, implying that dusty galaxies are preferentially the hosts
of high redshift SNe. 

The most likely reason for this preference is that high-z dusty galaxies are either in the throes of a starburst
or have undergone a large burst of star-formation in their past history. In the former case, the dust would presumably
be localized around star-forming regions while in the latter case, the dust would have been produced in the shells of the
AGB stars that have evolved out of the starburst. In either case, the evolving starburst would be ripe for the progenitors
of both types of SNe.

A possible alternative to this conclusion
is that the 24$\mu$m emission arises from an active galactic nucleus in the host galaxy.
AGN-dominated systems are thought to have L$_{\rm X}$/L$_{\rm IR}>0.01$ while starburst
dominated systems typically have L$_{\rm X}$/L$_{\rm IR}<10^{-3}$.
Matching the HDF-N 2Ms X-ray catalog and CDF-S 1Ms catalog (Alexander et al. 2003)
results in X-ray detections of six SN host galaxies. These are the
host galaxies of SN1997ff, SN2002dc, SN2003bb, SN2003er, SN2003es, SN2002kd. 
The hosts of 2003er, 2003bb, 2002dc and 2002kd are
extremely faint X-ray sources that are only detected in the soft band 
and whose X-ray flux is probably dominated by star-formation. The X-ray luminosities are in the range 10$^{40}$-10$^{41}$~erg~s$^{-1}$
implying L$_{\rm X}$/L$_{\rm IR}<<10^{-3}$, strongly indicative of star-formation dominated sources.
The hosts of SN 1997ff and 2003es are detected in the hard-band
with luminosities of $\sim$3$\times$10$^{42}$~erg~s$^{-1}$. The latter has a measured photon
index of $\Gamma\sim1.6$ indicative of AGN emission and has an L$_{\rm X}$/L$_{\rm IR}$ ratio of 0.05 which is
a factor of 20 higher than the former. This implies that the
X-ray and 24$\mu$m emission in the SN2003es host
is arising from a Seyfert nucleus while the distinction is unclear
for the host of SN1997ff. The remainder of the 24$\mu$m detected hosts are undetected in the X-rays with
the resultant constraints on the L$_{\rm X}$/L$_{\rm IR}$ ratio falling in the starburst regime.
As a result, we believe that for the majority of the detected SNe host galaxies,
the mid- and far-infrared emission cannot be powered by an AGN.

Another alternative which has been suggested for the far-infrared emission of the host galaxies is 
diffuse cirrus (Farrah et al. 2004). Clements et al. (2004, 2005) and Farrah et al. (2004) obtained deep submillimeter
observations of a sample of Type Ia SN hosts. Despite a low detection rate, they performed stacking analysis to
get a constraint on the total dust emission in the hosts. Based on the optical extinction value and
450$\mu$m to 850$\mu$m colors of one of the hosts,
they interpret the source of far-infrared emission to be cirrus. Among our sample,
only the host galaxy of SN 2002dc is seen in the 850$\mu$m data with a flux density of 1.7$\pm$0.4 mJy (GN13 in 
Pope et al. 2005). This is perfectly consistent with the low detection rate seen in the aforementioned submillimeter surveys.
The predicted 24$\mu$m/850$\mu$m flux density ratio due to cirrus
emission in a galaxy at $z\sim0.5$ should be about 0.05 (e.g. Arendt et al. 1998)
while the observed ratio for the SN2002dc host is 0.22$\pm$0.05, indicating that
cirrus, whose spectrum peaks at $\sim$100$\mu$m cannot be the source of emission in this host.
If cirrus were to dominate the mid-infrared emission, then the submillimeter flux would be exceeded
ruling out this possibility.
We also utilize the template SED which was used to derive the bolometric correction, 
to derive an 850$\mu$m flux density for each host galaxy. We find that the typical 850$\mu$m
flux density is $\sim$0.1mJy while the 1$\sigma$ uncertainty in the 850$\mu$m data is the range 0.5-1.5 mJy.
The stacked average 850$\mu$m
flux density for the 26 host galaxies (excluding 2002dc) yields -0.27$\pm$0.20 mJy which is 
consistent with the average value of 0.1 mJy derived from the templates.

Significant efforts have been made to generate Type Ia SNe Hubble diagrams partitioned by host galaxy 
morphological type (Sullivan et al. 2003) to mitigate the effects of dust. As a result it is useful to compare the
24$\mu$m detections with the rest-frame optical/ultraviolet morphological type of the galaxies.
Visual inspection of the
ACS data indicates that the morphologies of the SNe host galaxies span the whole range including
disturbed ellipticals/S0 (SN1997ff, SN2003az, SN2003er, SN2003es, SN2003ew, SN2002lg), 
spirals (SN2002dc, SN2002kh, SN2003ba, SN2003bb, SN2003be, SN2003dx, SN2003eq, 
SN2002fz, SN2002hq, SN2002kb, SN2002kc, SN2002kd) and irregulars (SN1997fg, SN2002dd,
SN2003N, SN2003eb, SN2003et, SN2002fw). Of the Type Ia SNe hosts that are detected at 24$\mu$m,
irregulars make up 4/14, Early type S0/compact hosts make up 4/14 while spirals/disks make up 6/14, a distribution
quite similar to that compiled by Riess et al. (2002) for Type Ia SNe from the Asiago catalog (Cappellaro et al. 1997).
For the CC SNe hosts detected at 24\um, irregulars make up 2/13, early type S0/compact hosts 3/13, and spirals
8/13. Within the uncertainties, the morphological distribution of galaxies for the two types of SNe are not dissimilar.
Furthermore, the mid-infrared flux from the early-type galaxes exceeds their optical/near-infrared flux by factors
of $1-30$, comparable to the ratio in late-type galaxies. This is more than an order of 
magnitude above the level expected from stellar photospheric emission 
in dust-poor early type galaxies in the local Universe 
implying that in the high redshift Universe, significant quantities of
dust are present even in early-type galaxies (see also e.g. Krause et al. 2003). 

\section{Implications for SNe}

Of the 45 SNe in the GOODS fields which have hosts and SN types determined, 27 are Type Ia's while 18 are CC. At least 10 
of the 27 Type Ia host galaxies are detected at 24$\mu$m with significant, but blended detections of 4 others, for a detection
rate of 52\%.
In comparison, at least 13 (76\%) of the 18 CC host galaxies are detected at 24$\mu$m with one 
additional possibly blended detection. 
In all, the detected Type Ia SNe host galaxies include 9/18 Type Ia SNe that were
used by Riess et al. (2004) and Riess et al. (2003) to trace cosmological parameters. 

CC SNe are intrinsically 1-2 mags less luminous than Type Ia SNe. They are detected only at $z<1$ while
Type Ia SNe have been detected out to $z\sim1.8$. Since the bolometric correction in translating the observed mid-infrared
flux density of the host galaxies
to L$_{\rm IR}$ is strongly redshift dependent and the ACS data trace rest-frame optical colors at 
$z<1$ we restrict our comparison between the properties of
Type Ia and CC SNe hosts to a common redshift range i.e. $0.1<z<1$.

Despite the small sample, Type Ia SNe host galaxies (including the limits), 
have a median L$_{\rm IR}$ of 4$\times$10$^{10}$~L$_{\sun}$, which is very similar
to the value observed for the CC SNe hosts. None of the host galaxies would be classified
as ultraluminous infrared galaxies (ULIGs; L$_{\rm IR}>10^{12}$~L$_{\sun}$). 
A K-S test between the infrared luminosities of the two host
galaxy populations returns a probability of 0.5, suggesting that they are different subsets of the
field galaxy population.
The similarity in L$_{\rm IR}$ values is surprising because core-collapse SNe are thought
to be the end states in the evolution of very massive stars and would preferentially be associated with ongoing
star-formation.  Since it has been shown that the majority of the star-formation in the high redshift 
Universe takes place in
LIGs with 10$^{12}>$L$_{\rm IR}>10^{11}$~L$_{\sun}$ (Elbaz et al. 2002, Le Floch et al. 2005), CC SNe should prefentially 
occur in such galaxies. However, the observations suggest that Type Ia SNe, which are thought to have a long evolutionary
delay after the epoch of star-formation, occur in infrared luminous galaxies as well. 

A possible explanation for the absence of a strong association between infrared luminous
hosts and CC SNe is a selection effect due to obscuration of the SNe.
Infrared luminous galaxies in the local Universe appear to be undergoing heavily obscured starbursts with
gas column densities corresponding to A$_{\rm V}$ greater than a few magnitudes. CC SNe are fainter
than the Type Ia's (Dahlen \& Fransson 1998) and are at the limits of detectability
in the deepest optical surveys such as GOODS. The absence of a strong association between the detected
core-collapse SNe and LIGs/ULIGs suggests that a large fraction of core-collapse SNe are obscured which the 
observational selection effects make difficult to overcome. As a result, high-z core-collapse SN rates that have
currently been determined are largely underestimated which is also suggested by the analysis of Dahlen et al. (2004).

An intriguing problem is harmonizing the absence of red $B-V$ colors in the light curve of the 
Type Ia SNe (Riess et al. 2004, Perlmutter et al. 1999) with the 24$\mu$m detection of the host galaxies.
Although higher redshift supernovae show a factor of $\sim$2-3 larger scatter in their $B-V$ colors at peak
brightness, the average $B-V$ color is still consistent with no-evolution from low redshift. 
The possibilities to explain this are: 
\begin{itemize}
\item The dust is grey which does not result in differential reddening between the rest-frame B and V-bands but would
systematically dim the supernovae; In this case, the scatter is due to heterogeneity in the intrinsic properties
of the SN progenitor.
\item The supernova is located at a large spatial offset from the dusty environments which are responsible for the
infrared emission; 
\item Evolution of the color in the SN light curve due to progenitor evolution is serendipitously nullified by 
dust reddening.
\item Introduction of measurement uncertainties since the light curves of the high-z supernovae are standardized based on sparse sampling
of the light curve. In addition, an uncertain
$k-$correction is required to transform Type Ia SN light curves from the 
observed $i$ and $z$-band to rest-frame $B-V$ colors.
\end{itemize}
\noindent We can only evaluate the second of these hypotheses since we have the positions of the SNe within the
host galaxies, the resolved colors of the hosts and their infrared luminosities. 
Such an
analysis of a more limited sample of SNe has been performed in the submillimeter
previously by Farrah et al. (2004) and Farrah et al. (2002).
It would also be
useful to compare the $B-V$ colors of the GOODS Type Ia SNe at peak brightness with the detection
of 24$\mu$m flux density of the host galaxy but the colors of the SNe have not yet been published.

Figure 3 shows the spatial offset of the supernovae from the center of their host galaxies as a function
of the 24$\mu$m flux density of the host. We used the GALFIT software (Peng et al. 2002) to fit
two-dimensional S\'ersic (1968) surface brightness models to the ACS $z-$ band images of the host galaxies
(Ravindranath et al. 2004). GALFIT measures the S\'ersic index, which characterizes the shape of the light
profile, and the radius (r$_{\rm eff}$) enclosing 50\% of the total light.
Of the mid-infrared detected hosts,
we find that 10/14 of the Type Ia SNe and 9/13 of the CC SNe, are located within the half-light radius of the 
host galaxy. The median half-light radius (0.5$\arcsec$) at the median redshift (0.75)
corresponds to projected distances of $\sim$4~kpc in the standard 
cosmology, which is substantially larger
than the typical nuclear starbursts. If the 24$\mu$m emission arises in a nuclear starburst, as in many local
infrared luminous galaxies, then the dust may not be relevant for the SN colors. 

We can use the resolved ACS images of the host galaxies
to study the variation of observed $i_{p}-z_{p}$ pixel colors 
within the half-light radius, which could reveal the internal distribution of dust (e.g. Hatano et al. 1998, Papovich et al. 2003). 
For example, if 
the host galaxies had patchy/nuclear regions of star-formation which are dominating the mid and far-infrared
emission, then the internal colors of the galaxy would show a larger scatter than if the dust distribution were smooth.
The SNe host galaxies span a wide range of redshifts and we do not have high resolution data
in all passbands to sample the same rest-frame colors of all galaxies. Selecting the reddest passbands ($i$ and $z$)
from the GOODS ACS data and only the SNe hosts at $z<1$, provides a measure of the rest-frame optical colors of the galaxies. 
By limiting the color measurement to within the half-light radius, we can study the highest S/N area where the color
is not affected by the background noise.
Figure 4 illustrates
the average global color ($i-z$) of the host galaxy and root mean square (rms) variation in the 
pixel $i_{p}-z_{p}$ colors of the SN 
host galaxies plotted against the infrared luminosity.
We find that neither the average nor the variation of colors within the host galaxy show any trend with L$_{\rm IR}$. 
The global colors as well as the scatter in internal colors
of the Type Ia SNe hosts and CC SNe hosts are indistinguishable from each other with a K-S probability of 0.8.
Host galaxies that are detected at 24$\mu$m have significantly redder colors
($<i-z>$=0.39 mag) than those of the undetected hosts ($<i-z>$=0.28 mag).
Rather surprisingly, the patchiness of the dust distribution as measured by the rms of the pixel
$i_{p}-z_{p}$ colors within the half-light radius is 0.1 mag larger for the 24$\mu$m undetected hosts.
The lack of a strong trend between the rest-frame optical/UV colors and the mid-infrared detection implies
that the rest-frame optical/UV colors are only a weak indicator of dust distribution within galaxies.
However, the fact that the 24\um\ detected galaxies have redder internal colors but smaller scatter
suggest that the dust that is responsible
for the mid- and far-infrared emission has a smooth distribution.

\section{Conclusions}

We present deep mid- and far-infrared observations of SNe host galaxies in the GOODS fields. About 60\% of the hosts are
detected indicating substantial quantities of dust in SNe hosts. The detections span a range of galaxy morphological
types, including early-type galaxies. 
The high detection rate implies that high redshift SNe preferentially occur
in dusty galaxies because these objects have undergone a starburst in their recent past which would provide
a larger population of SN progenitors. The median derived far-infrared luminosities
of these galaxies is about 3$\times$10$^{10}$~L$_{\sun}$. This is similar to that
of typical dusty star-forming field galaxies. 
The median bolometric luminosity of
Type Ia SNe hosts is very similar to that
of core-collapse SNe which is surprising because the latter are
thought to preferentially occur in star-forming galaxies.
We interpret this as a selection effect since dust obscuration in infrared luminous galaxies
would prevent the detection of SNe, particularly
the fainter core-collapse SNe, at high redshifts. 
This indicates that a large fraction of core-collapse
SNe at $z\sim0.5$ are undetected in deep optical surveys. 

The mean and scatter in pixel colors within the hosts is used to constrain the distribution of
dust within the galaxies. 
24$\mu$m hosts are redder by $\sim$0.1 mag which is
inferred to be due to a smooth distribution of dust, translating to an average A$_{\rm V}$ of 0.5 mag.
The majority of SNe are located within the half-light radius of their host galaxies where the effect
of a smooth dust distribution would be more pronounced. This cannot be reconciled with the absence of
reddening in the rest-frame $B-V$ colors of Type Ia SNe. 
Unfortunately, the SN colors are not measured accurately enough at high redshift
to determine the presence of dust along the SN line of
sight and optical to near-infrared colors of high-redshift SNe should be measured in
future missions which intend to target supernovae as precision cosmological tools.

\acknowledgements
Support for this work, part of the Spitzer Space Telescope Legacy Science Program, was provided by NASA through an award
issued to the Jet Propulsion Laboratory, California Institute of Technology under NASA contract 1407. We wish to
acknowledge Colin Borys for extensive work on the SCUBA supermap in the HDF-N and
Lexi Moustakas for developing IDL routines to access the GOODS Hubble/ACS images.
We also acknowledge the referee for a careful reading of the manuscript and suggested improvements.

\clearpage

\begin{figure}
\epsscale{0.8}
\plotone{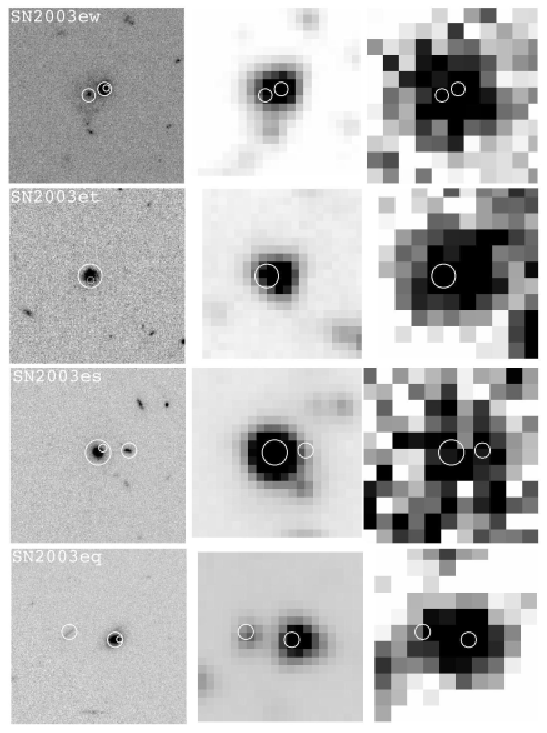}
\caption{
12$\arcsec$ snap shot images of Spitzer counterparts to Hubble-detected SNe host galaxies. From
left to right are the astrometrically aligned ACS $z$-band, IRAC 3.6$\mu$m and MIPS 24$\mu$m images. North is up,
East to the left. Small circles within the host indicate the position of the SN within
the SN host galaxy. Nearby galaxies are also circled to illustrate the possibility of blending of the host galaxy flux.
In certain cases (e.g. SN2003ew), the emission at Spitzer wavelengths is blended even at the IRAC
resolution and we are unable to disentangle the 24$\mu$m flux densities in those cases.
}
\end{figure}

\begin{figure}
\plotone{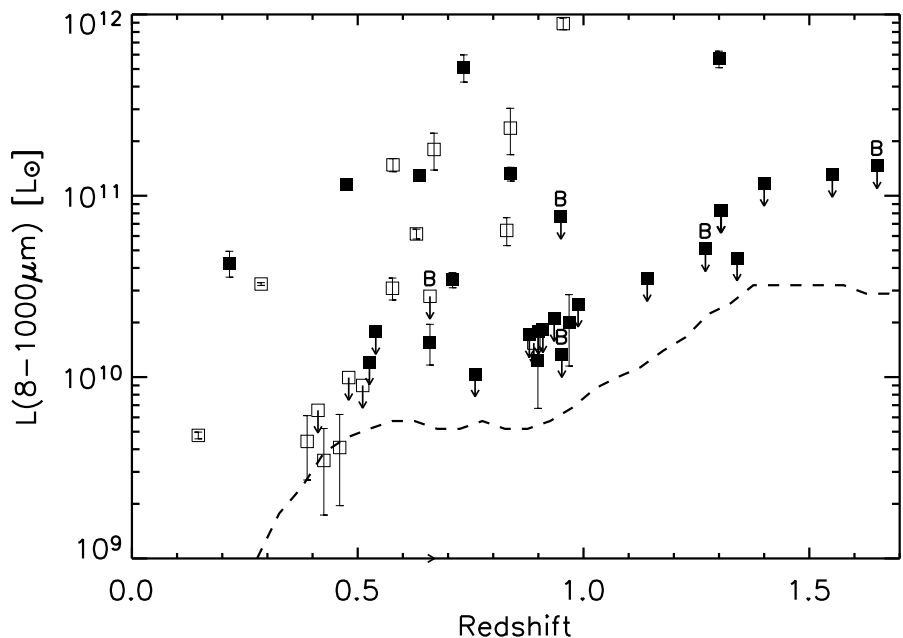}
\caption{Bolometric/far-infrared luminosities of SNe host galaxies derived from the observed
24$\mu$m flux density. Solid squares
are Type Ia and empty squares are Type II/CC. 
The dashed line indicates the L$_{\rm IR}$ value for a 24$\mu$m flux density of 10$\mu$Jy.
Blended host galaxies, labelled B above the data point, have upper limits to their L$_{\rm IR}$ based on the total 24$\mu$m 
flux density at that position.
Undetected host galaxies are shown by the L$_{\rm IR}$ corresponding to the 3$\sigma$ upper limit to their 24$\mu$m
flux density.  Rather surprisingly, Type Ia SNe host
galaxies have far-infrared luminosities which are similar to those of Type II SNe hosts.
Uncertainties in the derived L$_{\rm IR}$ show the quadrature sum of the
uncertainties due to systematics in the template and 24$\mu$m signal to noise.
}
\end{figure}
\begin{figure}
\plotone{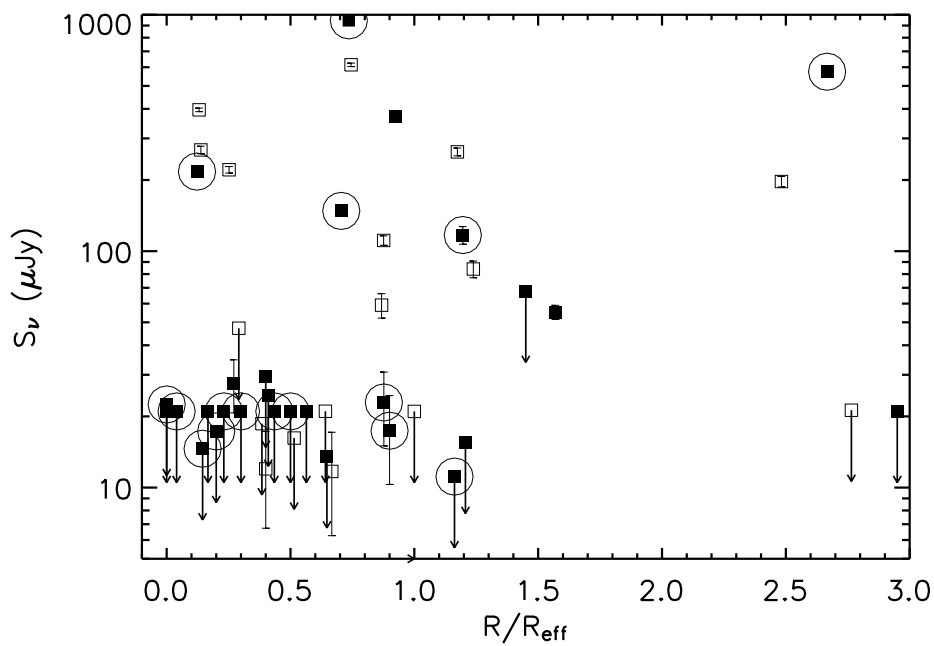}
\caption{The 24$\mu$m flux density of SNe host galaxies plotted against the ratio of their spatial offset from the
nucleus of the galaxy to the ACS $z-$band half-light radius. Type Ia SNe are shown as filled squares,
CC SNe as empty squares. Empty circles denote Type Ia SNe which have been used to probe cosmological parameters
in Riess et al. (2004). The majority of SNe are located within the half-light radius of the 
host galaxies where effects such as presence of dust extinction might be important. 
SNe are detected even in galaxies with large 24$\mu$m flux density suggesting that there are no
obvious selection effects against detecting mildly obscured SNe even at high-z.
}
\end{figure}
\begin{figure}
\plotone{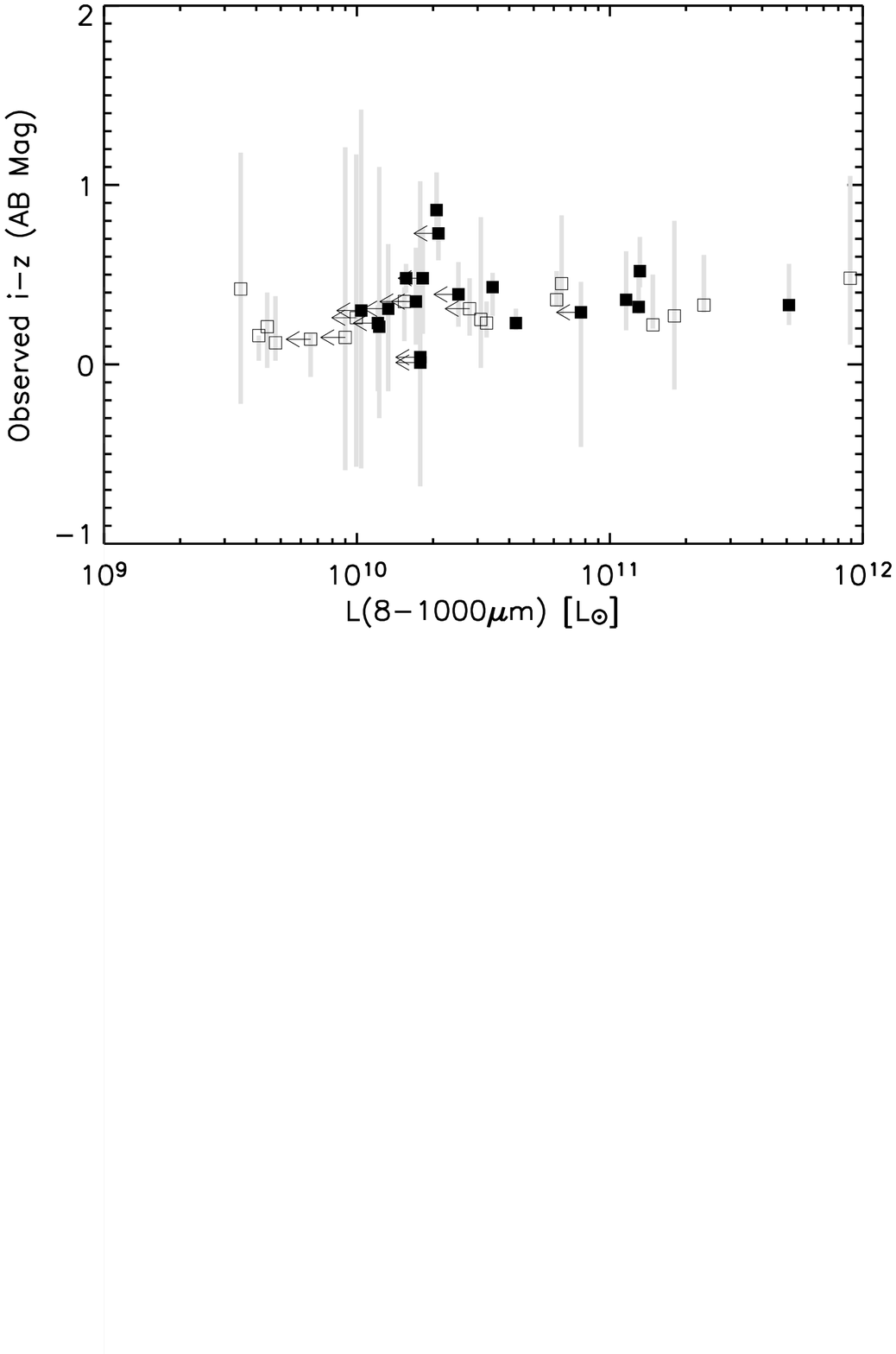}
\caption{
The average colors of SNe host galaxies (filled and empty symbols representing Type Ia and CC SNe respectively)
with the grey lines showing the range of
variation in the internal $i_{p}-z_{p}$ colors of pixels which are within the half-light 
radius of the SNe host galaxy. Only SNe in the 0.1$<$z$<$1 range are considered here, so
that the $i-z$ colors trace rest-frame optical colors of the hosts.
There is no trend between the bolometric luminosity of the hosts
and the color scatter which would be the case if the far-infrared luminosity were coming from a nuclear starburst.
However, 24$\mu$m detected galaxies are on average, redder by $\sim$0.1 mag than 24$\mu$m undetected hosts.
}
\end{figure}

\begin{deluxetable}{ccccccccc}
\tabletypesize{\scriptsize}
\tablecaption{Mid-infrared emission from SNe Host Galaxies in the GOODS-North field}
\tablewidth{0pt}
\tablehead{
\colhead{Name} &
\colhead{SN RA} &
\colhead{SN DEC\tablenotemark{c}} &
\colhead{z} &
\colhead{Type} &
\colhead{Host RA} &
\colhead{Host DEC\tablenotemark{c}} &
\colhead{S$_{\nu}$(24)} &
\colhead{S$_{\nu,{\rm err}}$(24)} \\
\colhead{} &
\multicolumn{2}{c}{(J2000)} &
\colhead{} &
\colhead{} &
\multicolumn{2}{c}{(J2000)} &
\multicolumn{2}{c}{$\mu$Jy}
}
\startdata
1997ff 			& 189.18378 & 62.21242 & 1.65 & Ia & 189.18381 & 62.21245 & $<$2.46E+01 & 4.51E+00\\ 
1997fg 			& 189.24039 & 62.22083 & 0.952 & Ia & 189.24030 & 62.22092 & $<$1.35E+01 & 8.34E+00 \\
2002dc\tablenotemark{b} & 189.20767 & 62.22028 & 0.475 & Ia & 189.20727 & 62.22030 & 3.71E+02 & 1.04E+01 \\
2002dd 			& 189.23068 & 62.21281 & 0.95 & Ia & 189.23107 & 62.21277 & $<$6.77E+01 & 7.09E+00 \\ 
2002kh 			& 189.06992 & 62.24371 & 0.71 & Ia & 189.07024 & 62.24361 & 5.52E+01 & 3.79E+00 \\ 
2002ki 			& 189.36822 & 62.34426 & 1.141 & Ia & 189.36822 & 62.34429 & ... & 5.76E+00 \\
2002kl 			& 189.45542 & 62.23493 & 0.412 & CC & 189.45532 & 62.23507 & ... & 7.08E+00 \\
2003N 			& 189.28858 & 62.18346 & 0.425 & CC & 189.28858 & 62.18340 & 1.20E+01 & 5.27E+00 \\
2003az 			& 189.33196 & 62.31032 & 1.27 & Ia & 189.33197 & 62.31035 & $<$1.46E+01 & 6.44E+00 \\
2003ba 			& 189.06617 & 62.21037 & 0.286 & CC & 189.06633 & 62.21040 & 3.96E+02 & 6.90E+00 \\
2003bb 			& 189.10196 & 62.14304 & 0.955 & CC & 189.10176 & 62.14349 & 6.14E+02 & 1.04E+01 \\
2003bc 			& 189.15858 & 62.16471 & 0.511 & CC & 189.15922 & 62.16484 & ... & 5.40E+00 \\
2003bd\tablenotemark{a} & 189.35442 & 62.22143 & 0.67 & Ia & ... & ... & ... & ... \\
2003be 			& 189.10821 & 62.11534 & 0.636 & Ia & 189.10832 & 62.11533 & 2.17E+02 & 6.69E+00 \\
2003dx 			& 189.13208 & 62.14676 & 0.46 & CC & 189.13200 & 62.14675 & 1.17E+01 & 5.44E+00 \\
2003dy 			& 189.28801 & 62.19120 & 1.340 & Ia & 189.28793 & 62.19129 & ... & 3.71E+00 \\
2003dz 			& 189.16629 & 62.13121 & 0.48 & CC & 189.16636 & 62.13117 & ... & 7.01E+00 \\
2003ea 			& 189.30017 & 62.21054 & 0.89 & CC & 189.30010 & 62.21055 & ... & 6.21E+00 \\
2003eb 			& 189.31303 & 62.22599 & 0.899 & Ia & 189.31289 & 62.22620 & 1.74E+01 & 7.09E+00 \\
2003en 			& 189.13812 & 62.22985 & 0.54 & Ia & 189.13810 & 62.22982 & ... & 9.86E+00 \\
2003eq 			& 189.45152 & 62.22641 & 0.839 & Ia & 189.45164 & 62.22638 & 1.48E+02 & 7.28E+00 \\
2003er 			& 189.13446 & 62.12634 & 0.63 & CC & 189.13492 & 62.12615 & 1.11E+02 & 5.13E+00 \\
2003es 			& 189.23079 & 62.21987 & 0.968 & Ia & 189.23103 & 62.21979 & 2.29E+01 & 7.90E+00 \\
2003et 			& 188.98279 & 62.22568 & 0.83 & CC & 188.98275 & 62.22577 & 8.40E+01 & 6.80E+00 \\
2003eu 			& 189.02458 & 62.18368 & 0.76 & Ia & 189.02456 & 62.18362 & ... & 5.18E+00 \\
2003ew 			& 189.11572 & 62.19018 & 0.66 & CC & 189.11586 & 62.19021 & $<$4.72E+01 & 6.04E+00 \\
2003lv 			& 189.37063 & 62.19095 & 0.935 & Ia & 189.37063 & 62.19095 & ... & 7.49E+00 \\
\enddata
\tablenotetext{a}{This SN has no detectable host galaxy at any wavelength in the optical-IR regime.}
\tablenotetext{b}{This SN host is the only one detected in the 850$\mu$m SCUBA HDF-N supermap with S$_{\rm 850}$=1.70$\pm$0.44 mJy.}
\tablenotetext{c}{The GOODS-N coordinate system is known to be offset from the world coordinate system by -0.38 arcsec in declination i.e. $\delta$WCS=$\delta$GOODS-N$-0.38\arcsec$. All coordinates presented here have been corrected for this offset.}

\end{deluxetable}

\begin{deluxetable}{ccccccccc}
\tabletypesize{\scriptsize}
\tablecaption{Mid-infrared emission from SNe Host Galaxies in the GOODS-South field}
\tablewidth{0pt}
\tablehead{
\colhead{Name} &
\colhead{SN RA} &
\colhead{SN DEC} &
\colhead{z} &
\colhead{Type} &
\colhead{Host RA} &
\colhead{Host DEC} &
\colhead{S$_{\nu}$(24)} &
\colhead{S$_{\nu,{\rm err}}$(24)} \\
\colhead{} &
\multicolumn{2}{c}{(J2000)} &
\colhead{} &
\colhead{} &
\multicolumn{2}{c}{(J2000)} &
\multicolumn{2}{c}{$\mu$Jy}
}
\startdata
1999gu & 53.25042 & -27.86111 & 0.147 & CC & 53.25078 & -27.86123 & 2.63E+02 & 1.00E+01\\
2002fv\tablenotemark{a} & 53.09471 & -27.85261 & ...   & CC & ...      & ...       &  ...     & 7.00E+00\\
2002fw & 53.15635 & -27.77956 & 1.300 & Ia & 53.15649 & -27.77962 & 1.17E+02 & 1.00E+01\\
2002fx & 53.02833 & -27.74289 & 1.400 & Ia & 53.02838 & -27.74289 &  ...     & 7.00E+00\\
2002fy & 53.07550 & -27.69878 & 0.880 & Ia & 53.07553 & -27.69877 &  ...     & 7.00E+00\\
2002fz & 53.20236 & -27.90493 & 0.838 & CC & 53.20200 & -27.90447 & 1.97E+02 & 1.00E+01\\
&&&&&&&                                                             1.73E+02\tablenotemark{b} & 2.66E+01\\
2002ga & 53.13598 & -27.88799 & 0.988 & Ia & 53.13592 & -27.88797 & ...      & 7.00E+00\\
2002hp & 53.10335 & -27.77164 & 1.305 & Ia & 53.10335 & -27.77165 & ...      & 7.00E+00\\
2002hq & 53.12487 & -27.72980 & 0.669 & CC & 53.12512 & -27.72985 & 2.21E+02 & 7.00E+00\\
&&&&&&&                                                             1.85E+02\tablenotemark{b} & 3.43E+01\\
2002hr & 53.09409 & -27.69779 & 0.526 & Ia & 53.09408 & -27.69780 & ...      & 7.00E+00\\
2002hs & 53.07746 & -27.80936 & 0.388 & CC & 53.07740 & -27.80870 & 2.05E+01 & 7.00E+00\\

2002ht & 53.03888 & -27.69157 & 0.900 & Ia & 53.03878 & -27.69170 & ...      & 7.00E+00\\
2002kb & 53.17675 & -27.84039 & 0.578 & CC & 53.17680 & -27.84032 & 2.68E+02 & 1.00E+01\\
&&&&&&&                                                             2.13E+02\tablenotemark{b} & 2.63E+01\\
2002kc & 53.14467 & -27.66619 & 0.216 & Ia & 53.14475 & -27.66596 & 9.47E+02 & 1.00E+01\\
&&&&&&&                                                             9.20E+02\tablenotemark{b} & 4.04E+01\\
2002kd & 53.09308 & -27.74081 & 0.735 & Ia & 53.09413 & -27.74051 & 5.74E+02 & 1.00E+01\\
&&&&&&&                                                             3.67E+02\tablenotemark{b} & 3.03E+01\\
2002ke & 52.99488 & -27.75019 & 0.577 & CC & 52.99449 & -27.75010 & 5.91E+01 & 7.00E+00\\
2002lg & 53.14904 & -27.79967 & 0.660 & Ia & 53.14896 & -27.79969 & 2.77E+01 & 7.00E+00\\
2003aj & 53.18478 & -27.91848 & 1.304 & Ia & 53.18479 & -27.91846 & ...      & 7.00E+00\\
2003ak & 53.19546 & -27.91374 & 1.551 & Ia & 53.19538 & -27.91363 & ...      & 7.00E+00\\
2003al & 53.02250 & -27.74144 & 0.910 & Ia & 53.02251 & -27.74146 & ...      & 7.00E+00\\
2003lt & 53.17838 & -27.93142 & ...   & ... & 53.17833 & -27.93153 & ...     & 7.00E+00\\
2003lu & 53.15050 & -27.91706 & ...   & ... & 53.15036 & -27.91697 & 1.60E+01 & 7.00E+00\\
 2004R & 53.17208 & -27.77044 & ...   & ... & 53.17210 & -27.77040 & ...      & 7.00E+00\\
\enddata
\tablenotetext{a}{This SN has no detectable host galaxy at any wavelength in the optical-IR regime.}
\tablenotetext{b}{16$\mu$m flux densities of the host galaxies.}
\end{deluxetable}

\begin{deluxetable}{cccccccr}
\tabletypesize{\scriptsize}
\tablecaption{Observed and Derived Properties of SNe Hosts in GOODS-N}
\tablewidth{0pt}
\tablehead{
\colhead{SN Name} &
\colhead{Host {\it i}} &
\colhead{Host {\it z}} &
\colhead{r$_{\rm eff}$} &
\colhead{Separation} &
\colhead{$<i_{p}-z_{p}>$} & 
\colhead{$\sigma(i_{p}-z_{p})$} &
\colhead{L$_{\rm IR}$\tablenotemark{a}}\\
\colhead{} &
\multicolumn{2}{c}{mag} &
\colhead{$\arcsec$} &
\colhead{$\arcsec$} &
\multicolumn{2}{c}{mag} &
\colhead{L$_{\sun}$}
}
\startdata
1997ff & 24.99$\pm$0.11 & 24.04$\pm$0.07 & 0.39 & 0.16 & 1.00 & 0.72 & $<$1.44E+11\\
1997fg & 22.85$\pm$0.05 & 22.54$\pm$0.05 & 0.68 & 0.44 & 0.26 & 0.41 & $<$1.33E+10\\
2002dc & 21.63$\pm$0.05 & 21.27$\pm$0.05 & 0.91 & 0.84 & 0.41 & 0.22 & 1.51E+11\\
2002dd & 22.03$\pm$0.05 & 21.74$\pm$0.05 & 0.20 & 0.87 & 0.00 & 0.46 & $<$7.69E+10\\
2002kh & 20.80$\pm$0.05 & 20.37$\pm$0.05 & 0.79 & 1.24 & 0.39 & 0.12 & 3.44E+10\\
2002ki & 23.75$\pm$0.06 & 23.27$\pm$0.06 & 0.50 & 0.10 & 0.53 & 0.49 & $<$2.45E+10\\
2002kl & 22.87$\pm$0.05 & 22.73$\pm$0.05 & 0.30 & 0.47 & 0.05 & 0.12 & $<$6.56E+09\\
 2003N & 24.04$\pm$0.07 & 23.62$\pm$0.06 & 0.50 & 0.20 & 0.48 & 0.70 & 3.47E+09\\
2003az & 25.03$\pm$0.09 & 23.97$\pm$0.06 & 0.69 & 0.10 & 0.94 & 0.94 & $<$5.10E+10\\
2003ba & 19.48$\pm$0.05 & 19.25$\pm$0.05 & 1.53 & 0.20 & 0.25 & 0.10 & 3.26E+10\\
2003bb & 20.53$\pm$0.05 & 20.05$\pm$0.05 & 1.84 & 1.37 & 0.58 & 0.47 & 6.02E+11\\
2003bc & 21.24$\pm$0.05 & 21.09$\pm$0.05 & 2.39 & 1.23 & 0.31 & 0.90 & $<$8.99E+09\\
2003bd & ... 		& ... 		 & ...  & ...  & ...  & ...  & ...\\
2003be & 20.48$\pm$0.05 & 20.16$\pm$0.05 & 0.98 & 0.12 & 0.40 & 0.12 & 1.30E+11\\
2003dx & 22.72$\pm$0.05 & 22.56$\pm$0.05 & 0.27 & 0.18 & 0.09 & 0.07 & 4.09E+09\\
2003dy & 23.22$\pm$0.05 & 22.70$\pm$0.05 & 0.37 & 0.43 & 0.42 & 0.18 & $<$3.67E+10\\
2003dz & 24.69$\pm$0.08 & 24.43$\pm$0.08 & 0.39 & 0.25 & 0.30 & 0.87 & $<$9.94E+09\\
2003ea & 23.23$\pm$0.05 & 22.88$\pm$0.05 & 0.39 & 0.15 & 0.28 & 0.15 & $<$1.03E+10\\
2003eb & 22.74$\pm$0.05 & 22.53$\pm$0.05 & 1.00 & 0.90 & 0.40 & 0.70 & 1.23E+10\\
2003en & 24.53$\pm$0.06 & 24.49$\pm$0.06 & 0.30 & 0.12 & 0.13 & 0.28 & $<$1.78E+10\\
2003eq & 21.89$\pm$0.05 & 21.37$\pm$0.05 & 0.61 & 0.43 & 0.57 & 0.14 & 1.32E+11\\
2003er & 20.39$\pm$0.05 & 20.03$\pm$0.05 & 1.13 & 0.99 & 0.42 & 0.10 & 3.94E+10\\
2003es & 22.47$\pm$0.05 & 21.61$\pm$0.05 & 0.65 & 0.57 & 0.91 & 0.16 & 1.52E+10\\
2003et & 22.49$\pm$0.05 & 22.04$\pm$0.05 & 0.42 & 0.52 & 0.55 & 0.28 & 6.44E+10\\
2003eu & 24.14$\pm$0.08 & 23.84$\pm$0.08 & 0.63 & 0.76 & 0.42 & 1.00 & $<$9.19E+09\\
2003ew & 21.47$\pm$0.05 & 21.16$\pm$0.05 & 0.79 & 0.23 & 0.32 & 0.16 & $<$2.79E+10\\
2003lv & 22.48$\pm$0.05 & 21.75$\pm$0.05 & 0.41 & 0.00 & 0.72 & 0.14 & $<$1.38E+10\\
\enddata
\tablenotetext{a}{Infrared luminosity estimates for blended sources are the derived L$_{\rm IR}$ for the
host and companion galaxies as a whole. For sources undetected at 24$\mu$m, L$_{\rm IR}$ estimates are
derived from the 3$\sigma$ upper limit to the 24$\mu$m flux density.}
\end{deluxetable}

\begin{deluxetable}{cccccccr}
\tabletypesize{\scriptsize}
\tablecaption{Observed and Derived Properties of SNe Hosts in GOODS-S}
\tablewidth{0pt}
\tablehead{
\colhead{SN Name} &
\colhead{Host {\it i}} &
\colhead{Host {\it z}} &
\colhead{r$_{\rm eff}$} &
\colhead{Separation} &
\colhead{$<i_{p}-z_{p}>$} &
\colhead{$\sigma(i_{p}-z_{p})$} &
\colhead{L$_{\rm IR}$\tablenotemark{a}}\\
\colhead{} &
\multicolumn{2}{c}{mag} &
\colhead{$\arcsec$} &
\colhead{$\arcsec$} &
\multicolumn{2}{c}{mag} &
\colhead{L$_{\sun}$}
}
\startdata
1999gu & 18.57$\pm$0.05 & 18.45$\pm$0.05 & 2.07 & 2.43 & 0.20 & 0.18 & 4.77E+09 \\
2002fv & ... & ... & ...  & ... & ... & ... & ... \\
2002fw & 25.49$\pm$0.13 & 24.89$\pm$0.09 & 0.46 & 0.55 & 0.68 & 1.18 & 5.68E+11 \\
2002fx & 25.69$\pm$0.12 & 25.21$\pm$0.09 & 0.23 & 0.10 & 0.44 & 0.49 & $<$1.17E+11 \\
2002fy & 25.80$\pm$0.08 & 25.45$\pm$0.07 & 0.20 & 0.00 & 0.38 & 0.27 & $<$1.71E+10 \\
2002fz & 21.37$\pm$0.05 & 21.04$\pm$0.05 & 0.85 & 2.11 & 0.45 & 0.16 & 2.36E+11 \\
2002ga & 23.32$\pm$0.05 & 22.93$\pm$0.05 & 0.39 & 0.22 & 0.39 & 0.18 & $<$2.52E+10 \\
2002hp & 24.05$\pm$0.05 & 23.06$\pm$0.05 & 0.50 & 0.02 & 0.98 & 0.37 & $<$8.33E+10 \\
2002hq & 20.05$\pm$0.05 & 19.78$\pm$0.05 & 3.69 & 0.93 & 0.33 & 0.47 & 1.80E+11  \\
2002hr & 24.64$\pm$0.05 & 24.41$\pm$0.05 & 0.20 & 0.06 & 0.09 & 0.24 & $<$1.21E+10 \\
2002hs & 23.56$\pm$0.05 & 23.35$\pm$0.05 & 0.33 & 2.51 & 0.19 & 0.21 & 4.42E+09 \\
2002ht & 26.18$\pm$0.14 & 26.17$\pm$0.17 & 0.20  & 0.59 & 0.17 &  0.85 & $<$1.78E+10 \\
2002kb & 19.93$\pm$0.05 & 19.71$\pm$0.05 & 1.45 & 0.20 & 0.35 & 0.15 & 1.48E+11 \\
2002kc & 18.78$\pm$0.05 & 18.55$\pm$0.05 & 1.21 & 0.89 & 0.25 & 0.06 & 4.25E+10 \\
2002kd & 20.61$\pm$0.05 & 20.28$\pm$0.05 & 1.23 & 3.28 & 0.39 & 0.17 & 5.11E+11 \\
2002ke & 20.60$\pm$0.05 & 20.35$\pm$0.05 & 1.36 & 1.18 & 0.40 & 0.42 & 3.09E+10 \\
2002lg & 20.88$\pm$0.05 & 20.40$\pm$0.05 & 0.59 & 0.16 & 0.48 & 0.08 & 1.56E+10 \\
2003aj & 24.74$\pm$0.06 & 24.06$\pm$0.05 & 0.39 & 0.09 & 0.55 & 0.49 & $<$8.30E+10 \\
2003ak & 23.35$\pm$0.05 & 23.06$\pm$0.05 & 0.96 & 0.48 & 0.43 & 0.96 & $<$1.31E+11 \\
2003al & 23.45$\pm$0.05 & 22.97$\pm$0.05 & 0.48 & 0.08 & 0.47 & 0.30 & $<$1.82E+10 \\
2003lt & 24.58$\pm$0.08 & 24.17$\pm$0.08 & 0.30 & 0.37 & 0.28 & 0.72 & ... \\
2003lu & 20.10$\pm$0.05 & 19.91$\pm$0.05 & 1.55 & 0.57 & 0.21 & 0.16 & ... \\
 2004R & 24.17$\pm$0.05 & 23.84$\pm$0.05 & 0.32 & 0.20 & 0.23 & 0.20 & ... \\

\enddata
\tablenotetext{a}{Infrared luminosity estimates for blended sources are the derived L$_{\rm IR}$ for the
host and companion galaxies as a whole. For sources undetected at 24$\mu$m, L$_{\rm IR}$ estimates are
derived from the 3$\sigma$ upper limit to the 24$\mu$m flux density.}
\end{deluxetable}

\end{document}